\documentclass[prl, twocolumn, showpacs, amsmath, amssymb,superscriptaddress]{revtex4-1}
\usepackage{amssymb,amsfonts,amsmath} 
\usepackage{graphicx}
\usepackage{color}
\usepackage{hyperref}
\usepackage{longtable}
\usepackage{bbm}
\usepackage{ulem}
\usepackage{enumitem}
\usepackage{nicefrac}

\begin{document} 

\title{Floquet Engineering in Quantum Chains}

\author{D.M.\ Kennes}
\affiliation{Department of Physics, Columbia University, New York, NY 10027, USA}

\author{A. de la Torre}
\affiliation{Department of Physics, California Institute of Technology, Pasadena, CA 91125, USA}
\affiliation{Institute for Quantum Information and Matter, California Institute of Technology, Pasadena, CA 91125, USA}

\author{A. Ron}
\affiliation{Department of Physics, California Institute of Technology, Pasadena, CA 91125, USA}
\affiliation{Institute for Quantum Information and Matter, California Institute of Technology, Pasadena, CA 91125, USA}

\author{D. Hsieh}
\affiliation{Department of Physics, California Institute of Technology, Pasadena, CA 91125, USA}
\affiliation{Institute for Quantum Information and Matter, California Institute of Technology, Pasadena, CA 91125, USA}

\author{A.J.\ Millis}
\affiliation{Department of Physics, Columbia University, New York, NY 10027, USA}
\affiliation{Center for Computational Quantum Physics, The Flatiron Institute,  New York, NY 10010, USA}

\begin{abstract} 

We consider a one-dimensional interacting spinless fermion model, which displays the well-known Luttinger liquid (LL) to charge density wave (CDW) transition as a function of the ratio between the strength of the interaction, $U$, and the hopping, $J$. We subject this system to a spatially uniform drive which is ramped up over a finite time interval and becomes time-periodic in the long time limit. We show that by using a density matrix renormalization group (DMRG) approach formulated for infinite system sizes, we can access the large-time limit even when the drive induces finite heating. When both the initial and long-time states are in the gapless (LL) phase, the final state has power law correlations for all ramp speeds. However, when the initial and final state are gapped (CDW phase), we find a pseudothermal state with an effective temperature that depends on the ramp rate, both for the Magnus regime in which the drive frequency is very large compared to other scales in the system and in the opposite limit where the drive frequency is less than the gap. Remarkably, quantum defects (instantons) appear when the drive tunes the system through the quantum critical point, in a realization of the Kibble-Zurek mechanism.

%
\end{abstract}

\pacs{} 
\date{\today} 
\maketitle

The manipulation of materials properties by controlled application of high amplitude electromagnetic fields, with the ultimate goal of creating "quantum matter on demand", is attracting an increasing amount of attention\cite{Basov17,Mankowsky16rev}. As technology for generating intense electromagnetic pulses across a broad wavelength spectrum has become available, there is an urgency to understand how to use light to induce phenomena that are inaccessible in thermal equilibrium and study its interaction with complex phases of matter.     

A system exposed to a time periodic drive may be described by a "Floquet Hamiltonian"\cite{Shirley65} with a discrete time translation invariance. While Floquet systems have been studied extensively in atomic physics \cite{CohenTannouji98,Eckardt05,Zenesini09,Itin15,Meinert16}, less attention has been paid to solid state realizations due to the issue of runaway heating. If the drive period matches an excitation energy in the solid, then an ever increasing number of excitations may be generated,driving the system to the infinite temperature limit.
However, if the drive frequency is sufficiently detuned from simple excitation energies, for example by being very high \cite{Abanin15,Mori16,Kuwahara16} or being well within an excitation gap \cite{Mentink15,Claassen16,Peronaci17}, runaway heating only occurs at exponentially long times and there is a well defined intermediate time scale, typically referred to as the pre-thermal regime, where heating is negligible. 
The high frequency limit has the added simplification that a low order Magnus expansion \cite{Magnus64} can be employed to describe the driven system in terms of an effectively static Hamiltonian with renormalized parameters \cite{Bukov15,Bukov16}.

Most of the theoretical work on Floquet-like systems has been limited to qualitative analysis, (effectively) non-interacting models or small systems in the long-time limit. Important exceptions include the work of Poletti and Kollath \cite{Poletti11} where a one dimensional Bose-Hubbard model with a drive field ramped slowly up from zero was studied; that of Mentink, Balzer and Eckstein and Mendoza-Arenas {\it et al.} \cite{Mentink15,Mendoza-Arenas17} who performed dynamical mean field analyses of the destruction of antiferromagnetism upon application of time periodic fields, and its dependence on ramp speed \cite{Eurich11,Herrmann17,Schuler17}; and a general discussion of the Floquet adiabatic theory for different drive parameters \cite{Weinberg17}.

%

In this paper we present a comprehensive  study of an interacting quantum many-body model driven by electromagnetic radiation which vanishes at large negative times, is  periodic at large positive times,  and is ramped up at controllable rates.  We use a numerically exact density matrix renormalization group (DMRG)  method \cite{White92,Vidal07,Schollwock11}  that is formulated  in the thermodynamic (infinite system size) limit.\cite{Karrasch12a,Barther13,Kennes16} We show that this method allows us to access unprecedentedly long times in the pre-thermal regime where heating may be neglected. We are interested in the dependence of the properties of this pre-thermal states on the ramp speed of drive and frequency. We find that the long-time Floquet behavior can be qualitatively understood in terms of equilibrium models with renormalized parameters and a temperature which depends on ramp speed and other factors.\cite{Lorenzo17} 
Finally, we show that a key  feature of the  `Kibble-Zurek' case \cite{KZ,Manmana09,Gardas17}, in which the drive tunes the system across a quantum phase transition, is the appearance of quantal (instanton/anti-instanton) defects. Our results demonstrate the power of time-dependent DMRG to study Floquet engineering in interacting systems.

\begin{figure}[t]
\centering
\includegraphics[width=\columnwidth]{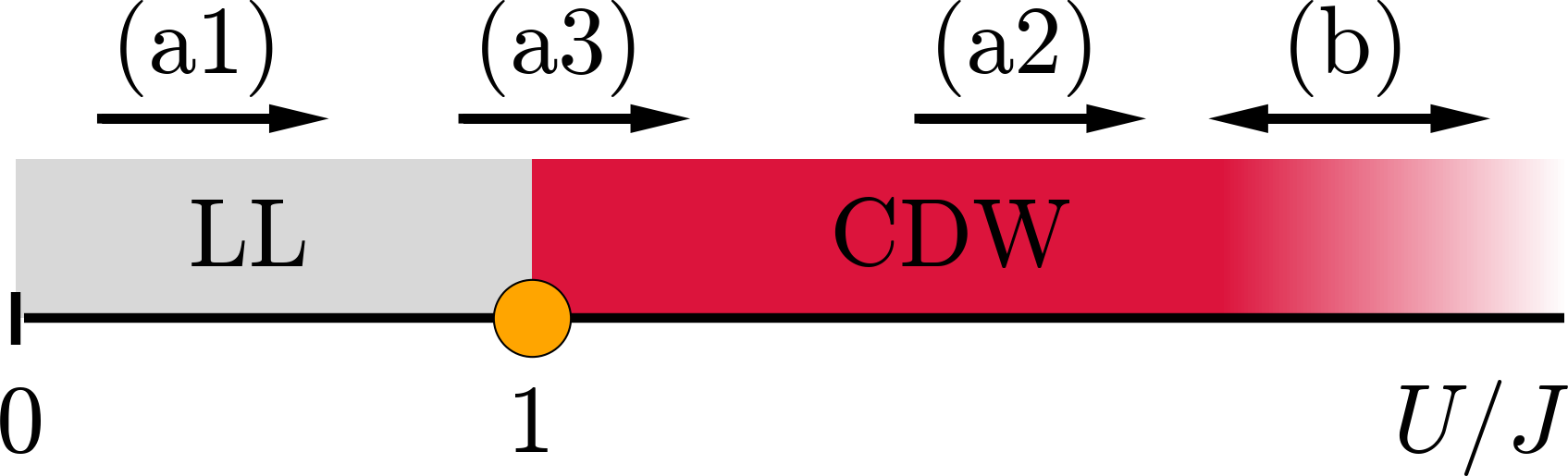}
\caption{Ground state phase diagram of Eq.~\ref{H} as a function of ratio of  interaction strength $U$ to hopping $J$  showing Luttinger Liquid  (LL), and charge density wave (CDW) phases.  Notations (a1-3) and (b) indicate 
the studied cases, with the direction in which the effective interaction strength can be tuned by a nonequilibrium drive indicated by the arrow head. The circular marker signals the location of the LL to CDW quantum phase transition.
}
\label{fig:plot1}
\end{figure}
We consider spinless fermions with a nearest neighbor interaction described by the following Hamiltonian
\begin{equation}
H(t)= \sum_{j} \left[-\frac{J(t)}{2}c_j^\dagger c_{j+1}+{\rm H.c.}
+U(n_j-\frac12)(n_{j+1}-\frac12)\right].
\label{H}
\end{equation}
The operators $c_i$ and $c^\dagger_i$ annihilate or create spinless fermions at site $i$ and $n_i=c^\dagger_ic_i$ is the site occupancy. We concentrate on the case of half filling,  with $U>0$ and  (without loss of generality) $J>0$. The equilibrium phase diagram is shown in Fig~\ref{fig:plot1}. We choose  $J(t)=Je^{i A(t)}$, with $A(t)$  the vector potential corresponding to a spatially uniform electric field $E=-\partial_tA$ and consider a harmonic drive with frequency $\Omega$ which is ramped on over a time interval $\tau$:  
\begin{equation}
A(t)=\frac{E_0}{\Omega}\sin(\Omega t)\left[\frac12 +\frac12 \tanh\left(\frac{t}{\tau}\right)\right]
\label{Adef}
\end{equation}
 
We consider two frequency regimes:  "Magnus", $\Omega\gg   J,U$ (we choose $\Omega=10J$) and "subgap",   $\Omega <\Delta$ (we choose $U=16J$ and $\Omega=0.6U$). Previous work on related models 
\cite{Abanin15,Mori16,Kuwahara16,Mentink15,Claassen16,Peronaci17} suggests that  in these  regimes a parametrically long intermediate time regime exists in which heating may be neglected and a steady state may be defined. As such, the long-time physics may be understood in terms of pseudo-equilibrium arguments based on Hamiltonians renormalized via an appropriate average over a drive period, effectively moving the system from one point to another  in the phase diagram (Fig.~\ref{fig:plot1}). In this language we distinguish the bare parameters (without the driving) from the effective ones (with the driving) such as the bare gap $\Delta$ and the effective gap $\Delta^{\rm eff}$ (both can be obtained by Bethe Ansatz using the values of $U$ and the bare hopping $J$ or effective hoping $J^{\rm eff}$, respectively).  In the Magnus case  it is argued \cite{Magnus64} that in the steady state one may simply replace $J(t)$ by its average over a period $2\pi/\Omega$, $J\to J^{\rm eff}=J_0(E_0/\Omega) J $. This leads to a decrease in the magnitude of $J$, because the Bessel function has magnitude less than $1$, i.e. an increase in the ratio $U/J$, implying that the drive moves the system to the right in Fig.~\ref{fig:plot1} as indicated by arrows in the cases (a1-3), either within the LL phase (a1), within the CDW phase (a2) or across the quantum critical point separating the two (a3). In the subgap regime, the modification of the  Hamiltonian parameters is more involved than in the Magnus case. As noted in Refs.~\cite{Mentink15,Claassen16}, if the drive period is small relative to the gap, analytical results may be obtained by retaining only processes that couple adjacent Floquet bands and averaging over a drive period. Applying this method to our model  we find that the long-time behavior may be described by the effective hopping 
\begin{equation}
J^{\rm eff}=J\sqrt{\sum\limits_{n=-\infty}^\infty \left(\frac{J_{n}\left(E_0/\Omega\right)}{1-n\Omega/U}\right)^2}
\label{eq:ana_pred}
\end{equation}
which may be either smaller or larger than the starting $J$ so the system may be moved either to the left or the right on the phase diagram of Fig.~\ref{fig:plot1} but of course only within the gapped phase (case (b)  Fig.~\ref{fig:plot1}).

We characterize the out of equilibrium behavior via the equal time density-density correlation function $C$, which depends on relative position $j$, ramp time $\tau$ and time, $t$:
\begin{equation}
C_{nn}(j,t,\tau)=\left\langle \left(n_0(t)-\frac12\right)\left( n_j(t)-\frac12\right) \right\rangle
\label{Cdef}
\end{equation}
focussing in particular on the $t$ and $\tau$ dependence of the large $j$ behavior.  In the Luttinger liquid phase at equilibrium as $T\rightarrow 0$, $C$ decays as a power law for large $j$ while in the CDW phase $C$ tends exponentially to a non-zero constant. At $T>0$ $C$ decays exponentially to zero at long scales, with exponent depending on phase, value of interaction, and temperature. 

We  use the DMRG methods of Refs.~\cite{Karrasch12a,Barther13,Kennes16}  to solve the model  (see supplemental information \cite{SI} for details). We start from the ground state corresponding to  $A(t)=0$ and integrate forward in time.   DMRG calculations  are limited by the growth of entanglement entropy; in the Magnus  and subgap  regimes the entanglement remains manageable because there is no runaway heating, allowing  us to reach large times.   In all cases except the `Kibble-Zurek' (a3) situation, we find (see the SI \cite{SI}) that after times $\sim 100J$ the system reaches a steady state, in which the properties (averaged over a few drive periods) become time-independent. We describe the steady state properties by comparing to a pseudo thermal state given by a diagonal density matrix. 

 
\begin{figure}[t]
\centering
\includegraphics[width=\columnwidth]{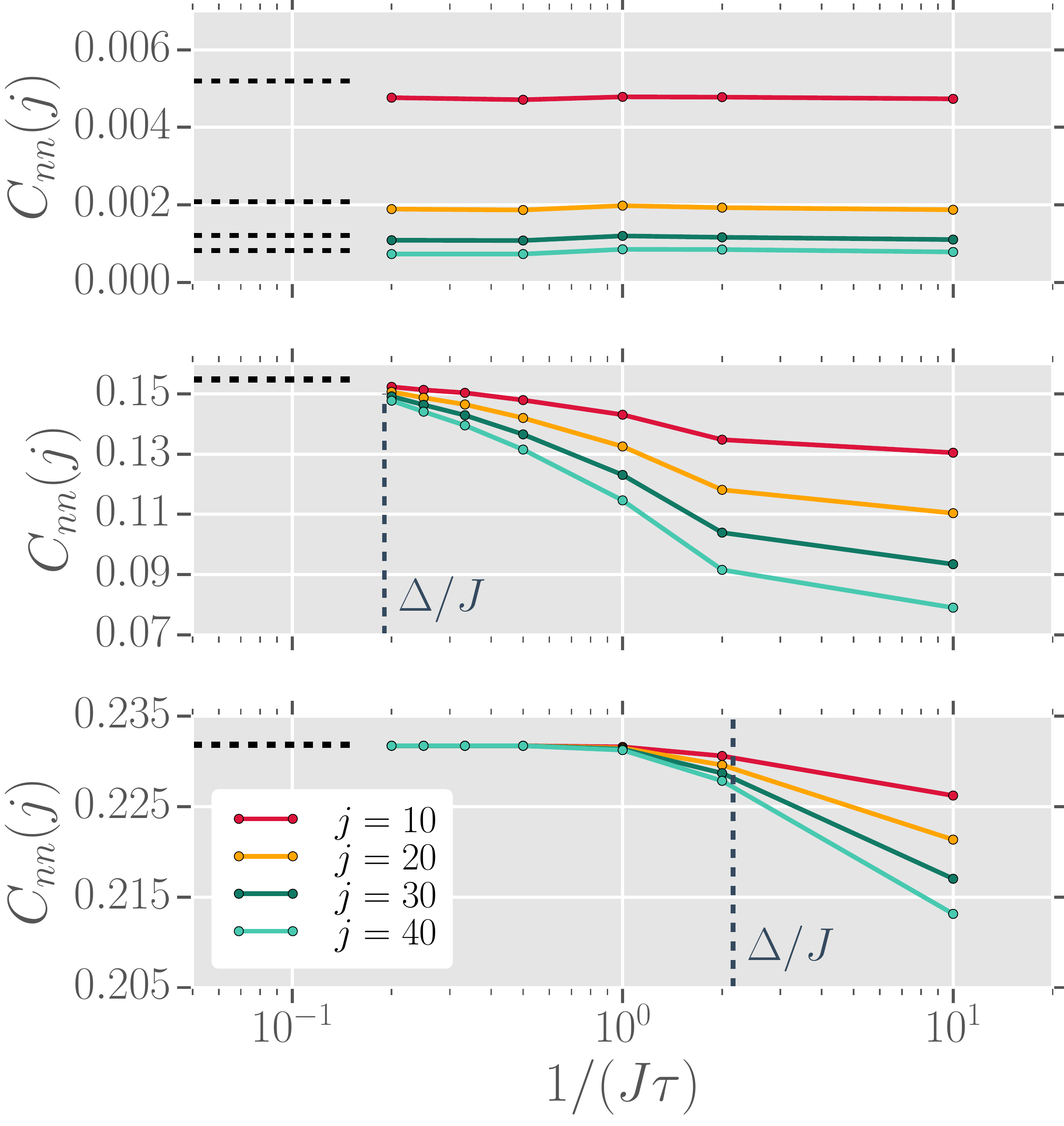}
\caption{Density-density correlation function $C(j)$ averaged over the time range $t=100/J$ to $t=400/J$ for different $j$ (solid lines) as function of inverse of ramp time $\tau$ for  $\Omega/J=10$, $E_0/\Omega=1$ and $T/J=0$ (Magnus limit). Top panel: initial correlation strength $U/J=0.5$ (LL phase); Magnus estimate of final correlation strength $U/J_{\rm eff}\approx0.65$. Middle panel: initial state $U/J=1.75$ (CDW), final $U/J_{\rm eff}\approx 2.29$ (CDW); bottom panel  initial state $U/J=4$ (CDW), final $U/J_{\rm eff}\approx 5.23$ (CDW). The ground state expectation values from the   effective Hamiltonian  ($J\to J^{\rm eff}$) are given as horizontal dashed lines and the gaps  $\Delta^{\rm eff}$ as vertical dashed lines.  }
\label{fig:plot_taudep}
\end{figure}


Fig.~\ref{fig:plot_taudep} shows the long-time behavior of $C$, as a function of inverse ramp time for different distances $j$.    The upper panel shows that when the system is in the Luttinger liquid phase both before and  after the ramp (case (a1)) the behavior is completely independent of the ramp time, and that values of the correlation functions are very close to those predicted by using the Magnus formalism to obtain an effective $J$ and then using equilibrium formulae to calculate the $T=0$ behavior.  As shown in the SI the exponent  characterizing the power-law decay is, within our numerical uncertainty, identical to Magnus estimate but the prefactor is slightly larger \cite{SI}.   This $T=0$-like Luttinger liquid behavior is also seen in the momentum dependences displayed in the SI \cite{SI}. Thus in this case the energy injected by a non-adiabatic ramp does not manifest itself as an effective temperature, even  for ramp time as low as $\tau=J/10$ (compare supplemental information Fig.~S2).  This finding is consistent with previous reports that the integrability of the system means that quenching of a LL from one time independent Hamiltonian to another preserves the basic power law decay \cite{Cazalilla06,Rentrop12,Kennes13,Mitra13}.  

The middle and lower panels of Fig.~\ref{fig:plot_taudep}  study two examples of the case (a2) where the perturbation is expected to shift the system from one point in the CDW regime to another.  We again find that after a transient period $\lesssim 100/J$ the system evolves to a steady state, but  in contrast to the LL to LL case we find strong dependence on the ramp time; note in particular the ramp speed-dependent exponential decays at large $j$. Comparison of the two cases indicates that  the time scale  governing the ramp speed dependence  is the inverse of the gap $\Delta^{\rm eff}$ of the final state (obtained from Bethe ansatz using  $J^{\rm eff}$ and $U$).
\begin{figure}[t]
\centering
\includegraphics[width=\columnwidth]{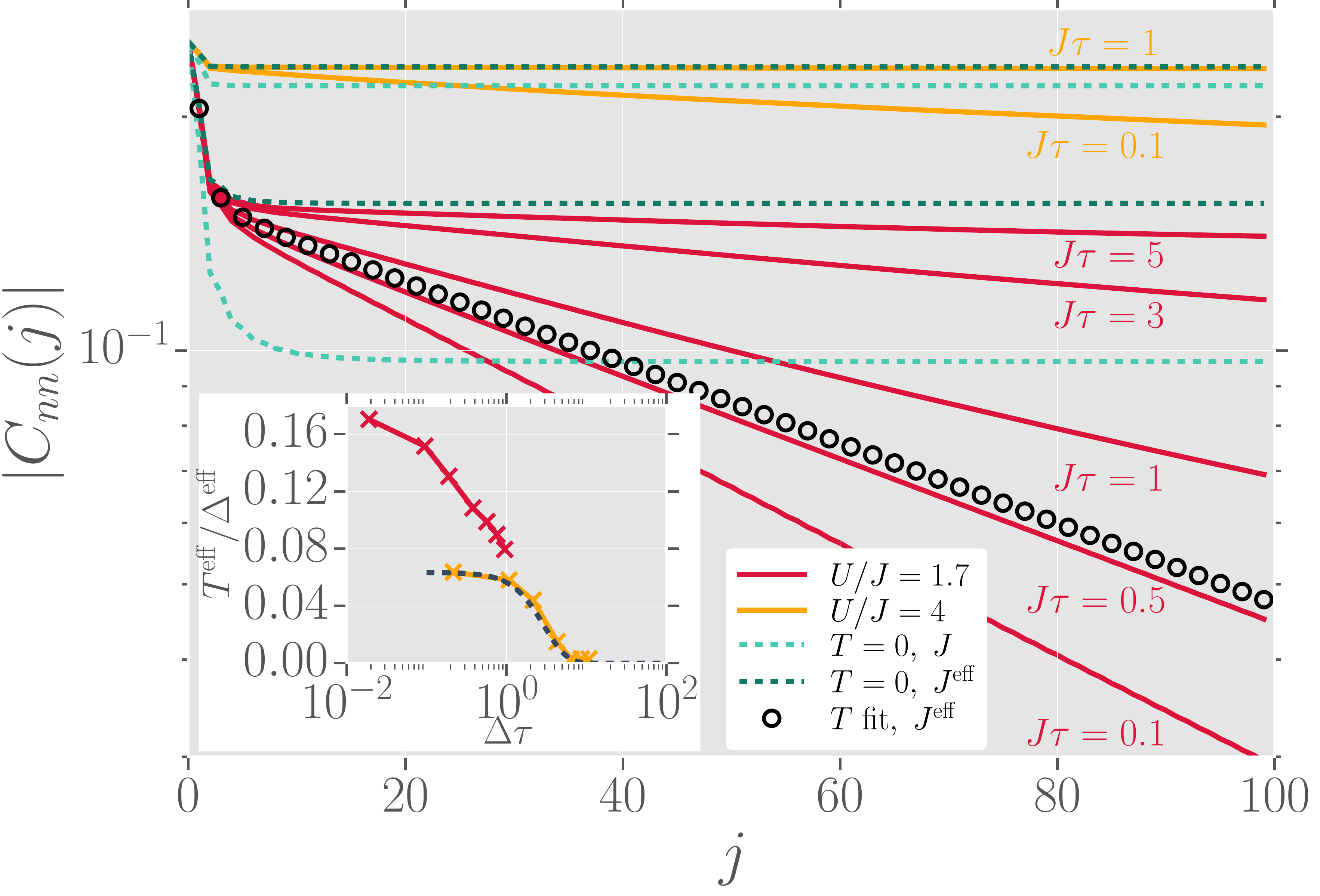}
\caption{ Main panel: Time-averaged density-density correlation function $C(j)$ (solid lines) for different $J\tau$ and $U/J$. 
The other parameters are  $\Omega/J=10$, $F_0/\Omega=1$ and $T/J=0$. The $T=0$ equilibrium prediction with $J\to J^{\rm eff}=J_0(E_0/\Omega) J $ (dark dashed lines) compares well to the time averaged results only in the limit of large $J \tau$. The light dashed lines are the undriven equilibrium results. Inset: Effective temperature (crosses) as function of $\tau$ extracted by fitting the slope of the exponential tail of the $U/J=1.7$ and $U/J=4$ lines in the main panel. One of these fits ($U/J=1.7$ and $J\tau=0.5$) of the slope of the exponential tail in $j$  is shown in the main panel by open circles. The approximate prediction $T^{\rm eff}/\Delta^{\rm eff}= a \Delta\tau\sinh\left(b \tau \Delta\right)$ is the dashed line (see SI \cite{SI}).   }
\label{fig:plot3}
\end{figure}

Fig.~\ref{fig:plot3} considers the case (a2) in more detail, plotting the $j$ dependence of the logarithm of $|C|$ for different ramp speeds and initial correlation strengths. Comparison to the equilibrium behavior suggests that  the energy put into the system by a rapid ramp produces an effective temperature $T_{\rm eff}$.  The inset shows the effective temperature (see supplementary information \cite{SI}) defined from  $C(j)=C_0e^{-\frac{\Delta^{\rm eff}}{T_{\rm eff}}}$ ($C_0$ is fitting constant). We see that for sufficiently adiabatic ramps, the effective temperature becomes unobservably small, but we believe that for all ramp speeds $T_{\rm eff}\neq 0$. Thus we argue that a non-adiabatic ramp creates a density of defects (as would also be created by a non-zero temperature) which are essentially randomly distributed and do not annihilate over the time scale of our simulations.   We finally note that although the long distance behavior is consistent with a nonzero temperature, the entire $j$ dependence cannot be described with a unique temperature/gap pair. As can be seen from the offset  between the open circles and the solid line,  the long distance decay is characterized by a prefactor different from the thermal equilibrium result. Relatedly,  the open circles agree very well with the short time behavior (see SI for more information \cite{SI}).

\begin{figure}[t]
\centering
\includegraphics[width=\columnwidth]{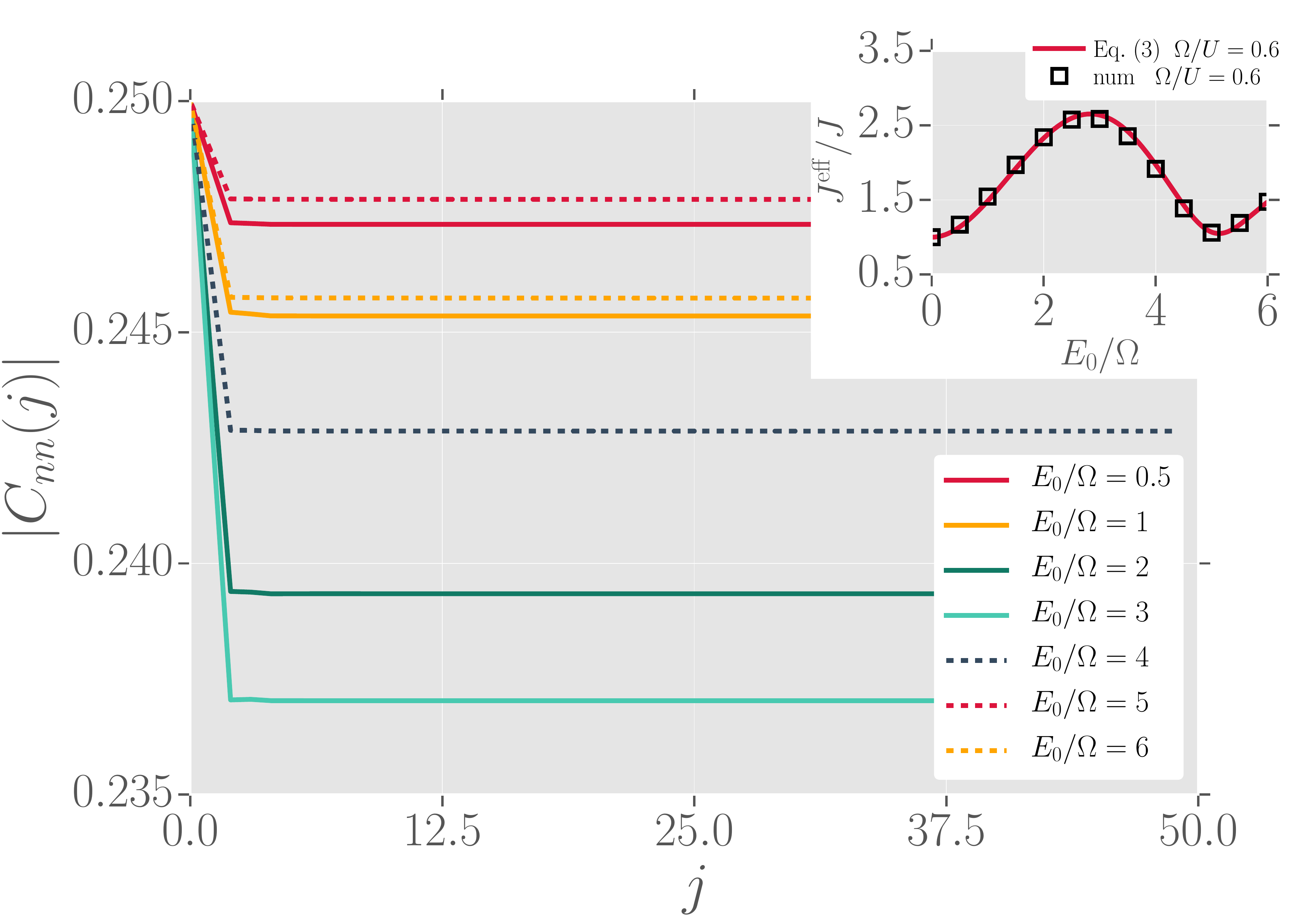}
\caption{Main Panel: Time-averaged density-density correlation function $C(j)$ (lines) for different $E_0/\Omega$. The other parameters are $J\tau=4$, $\Omega/U=0.6$, $U/J=16$ and $T/J=0$. Inset: comparison of Eq. ~\eqref{eq:ana_pred}  for effective hopping generated by sub-gap drive (solid line) and deduced from fits of the data shown in the main panel to the equilibrium $T-0$ formula for the long distance limit of $C$.  }
\label{fig:plot5}
\end{figure}

Very similar physics is obtained in the subgap regime (regime (b)). We find the same  dependence on ramp speed as in  case (a2). Thus, in Fig.~\ref{fig:plot5} we present only results in the quasi-adiabatic limit. For the case considered the long-time CDW amplitude is smaller than the initial amplitude (drive leads to weaker correlations), but with a non-monotonic dependence on the ratio of drive strength to frequency.  The inset confirms that the $J_{\rm eff}$ obtained by analysing the data in the main panel agrees perfectly with our theoretical prediction  Eq.~\eqref{eq:ana_pred}. Remarkably, equation~\eqref{eq:ana_pred} describes a highly tunable non-monotonic control of the ration $U/J$ either to larger or smaller values, which is beyond the control obtained in the Magnus  regime (see SI \cite{SI}).

\begin{figure}[t]
\centering
\includegraphics[width=\columnwidth]{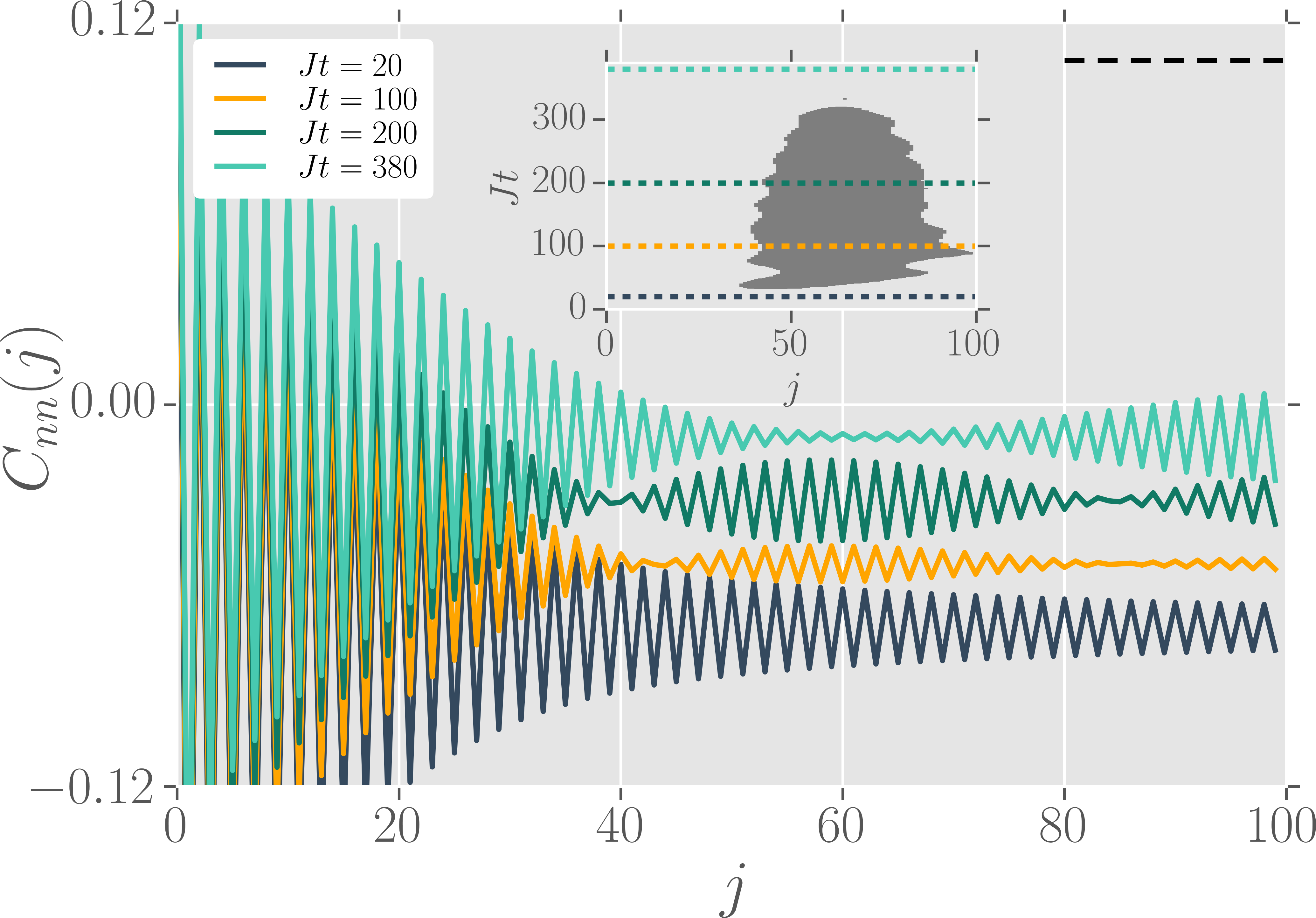}
\caption{
Main panel: The correlation function $C(j)$ at different times averaged over one drive period for the Kibble-Zurek (a3) case. Lines are shifted vertically for clarity of depiction; the midpoint of the oscillation is zero. The dashed black line gives the asymptotic value expected at large $j$ from a ground state calculation using the Magnus formalism, shifted to correspond to the longest time case.  Inset: time evolution of the phase  averaged over the drive period.  Light and dark gray denote the phase of $C$ $+1$ and $-1$ with respect to the small $j$ oscillation, respectively. The parameters are $U/J=0.95$, $E_0/\Omega=1.5$, $J\tau=4$ and $T/J=0$.   }
\label{fig:plot4_2}
\end{figure}

We finally show in Fig~\ref{fig:plot4_2} the case (a3) in which the drive tunes the system across the quantum critical point separating the LL and CDW phases.  The $Jt=20$ (lowest (black online)) curve is very similar to the short time behavior observed in the CDW to CDW quench (cases a2,b). where $C$ decreases with increasing $j$. This is qualitatively  consistent with a CDW-like phase with amplitude exponentially decaying at large distances.   But at slightly longer times ($Jt\sim 25$) a phase slip-anti phase slip pair  appears: as $j$ is increased the amplitude goes to zero, and then increases again but with the opposite phase (maxima in the positions where an extrapolation of the small $j$ curve would predict minima), then the amplitude again goes to zero and then the oscillations are in phase with the small $j$ ones. The phase slip and anti phase slip separate rapidly in space, then remain at a roughly fixed distance for a time interval $\sim 100/J$ and then re-coalesce  leaving a single phase regime (see inset). Such phase slip-anti phase-slip pairs were not observed in  any of our CDW $\to$ CDW cases (see SI \cite{SI}). 
Thus, we interpret the phase/anti-phase slip pairs as quantum defects produced {\it a la} Kibble-Zurek \cite{KZ} because the trajectory in parameter space passes close to the quantum critical point.


The first distance at which the phase slip-anti phase slip pair appears is somewhat dependent on ramp speed and drive strength, as is the time over which the phase slip and anti phase slip exist, but in all cases we have investigated the first time at which the pair appears is about the same ($Jt\sim 30$). The relatively weak dependence of many of the phase slip properties on parameters (see SI \cite{SI}) may be related to the logarithmic scaling associated with the Kosterlitz-Thouless-like criticality of the model at $U=J$. Note that unlike the defects which give rise to the exponential decay, these instantons anneal out in a finite time. 

In summary, this paper has established DMRG as an efficient tool to study Floquet engineering in interacting quantum systems in situations  where heating can be neglected over a wide time range. In this wide time range, it is generally accepted 
\cite{Abanin15,Mori16,Kuwahara16,Mentink15,Claassen16,Peronaci17} that the system is in a "pre-thermal" state described by a diagonal density matrix. we investigated three different Floquet engineering cases (LL $\to$ LL, CDW $\to$ CDW and LL $\to$ CDW), finding three different types of pre-thermal states. In the LL $\to$ LL case the energy put into the system as the drive is turned on does not manifest itself as an effective temperature. On the other hand in the CDW $\to$ CDW case the energy does lead to a behavior closely analogous to that found in thermal equilibrium, while the key feature of the LL $\to$ CDW case is  an interesting generation of finite lifetime quantal defects if the drive moves the system across a quantum phase transition. We also derived and numerically verified an expression for drive-induced parameter changes that goes beyond the standard Magnus expression and admits a weakening as well as a strengthening of the effective correlation parameter.  
Our work opens many directions for research. 

The methods presented here can be applied to many other one dimensional situations including ladders, higher-spin and longer ranged interaction spin chains and doped systems. This work sets the basis for the study of interacting spinful fermions of relevance  to quasi-1D conducting materials such as Li$_{0.9}$Mo$_6$O$_{17}$ (purple bronze) \cite{Schlenker85}, the organic salt TTF-TCNQ \cite{Jerome82}, or NbSe$_3$ \cite{He99}. NbSe$_3$ in particular  may be particularly amenable to Floquet engineering because its CDW gap scale is in the mid-infrared \cite{He99}, a region readily accessible by modern high pulse-energy lasers.

Other future directions include a study of the effect of pulses of finite duration.
On the analytic side, an improved understanding of the LL $\to$ CDW quench is urgently needed.


\begin{acknowledgments} 
\noindent \textit{Acknowledgments}---AJM and DMK were supported in part by the Basic Energy Sciences program of the Department of Energy under  grant DE-SC 0012375.  D.H. acknowledges support by the Army Research office MURI grant W911NF-16-1-0361. A.d.l.T.
acknowledges funding from the Swiss National Science Foundation through project P2GEP2\_165044.
Simulations were performed with computing resources granted by RWTH Aachen University under project rwth0013.
DMK also acknowledge the hospitality of the Center for Computational  Quantum Physics  of  the  Flatiron  Institute.
\end{acknowledgments}

{}

\clearpage\newpage

\centerline{\large{\bf Supplementary Information}}

\section{SI: Heating in Regime (a)}

\begin{figure}[b]
\centering
\includegraphics[width=\columnwidth]{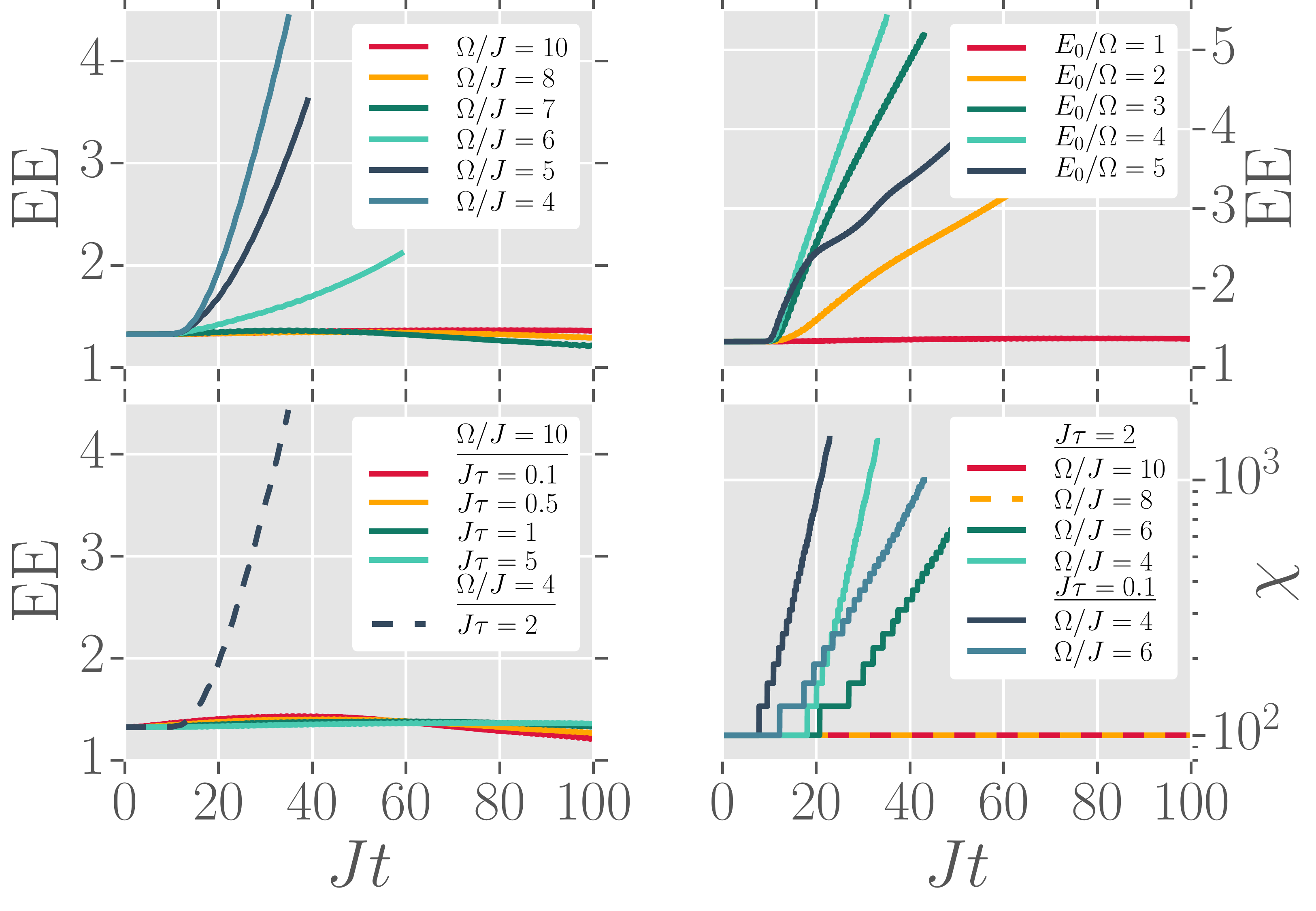}
\caption{ Entanglement entropy ($\rm EE$) for the periodically driven system in dependency of $\Omega$ (upper left), $E_0$ (upper right), $\tau$ (lower left) as well as the bond-dimension $\chi$ (keeping a fixed discarded weight of $10^{-8}$) in dependency of both $\tau$ and $\Omega$ (lower right). The other parameters are $U/J=0.5$, $\Omega/J=10$, $E_0/\Omega=1$, $J\tau=2$ and $T/J=0$ if not stated otherwise in the legends. }
\label{fig:plot_EE.pdf}
\end{figure}
Fig.~\ref{fig:plot_EE.pdf} shows the entanglement entropy defined by ${\rm EE}=-\rho_{\rm L} \log \left(\rho_{\rm L} \right)$, where $\rho_{\rm L}$ is the left half of our infinite system (upper left and right as well as lower left panel).  The lower right panel shows the bond-dimension $\chi$ (keeping a fixed discarded weight of $10^{-8}$). We concentrate on regime (a1) in which we tune a LL, but the other regimes look qualitatively similar. We find that entanglement growth (and with it heating) is suppressed entirely in the regime of high frequency and not to strong fields. In these regimes the value of $\tau$ has only a minor effect on the entanglement growth as can be seen by the parallel shifted curves in the lower right panel of Fig.~\ref{fig:plot_EE.pdf}.

\section{SI: Tuning a LL}
\begin{figure}[t]
\centering
\includegraphics[width=\columnwidth]{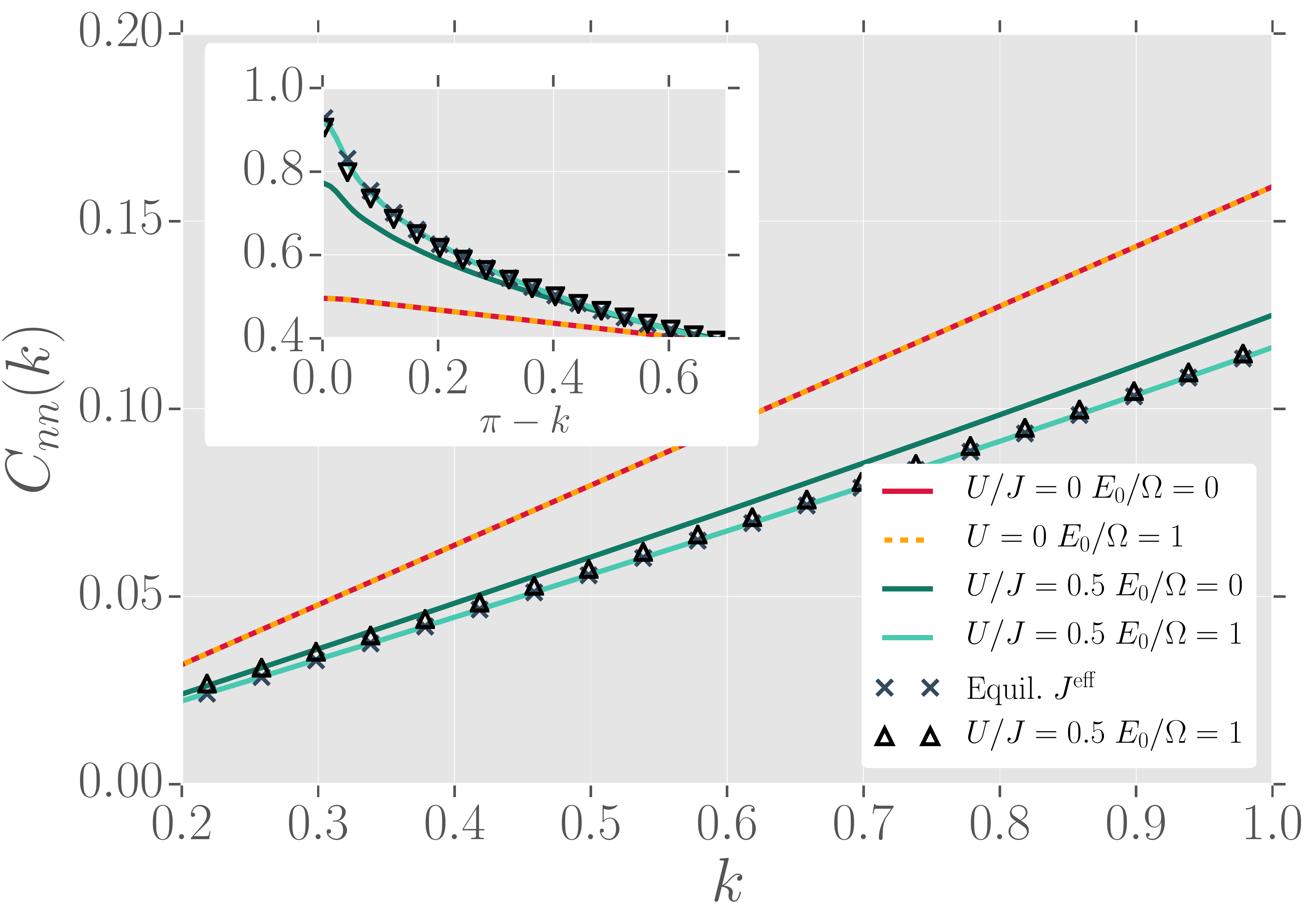}
\caption{Time-averaged density-density correlation function $C_{nn}(k)$ (lines and triangles) in momentum space. The other parameters are $J\tau=2$ (lines) and $J\tau=0.1$ (triangle), $\Omega/J=10$ and $T/J=0$. The $T=0$ equilibrium prediction for the effective (Magnus) Hamiltonian (crosses) with $J\to J^{\rm eff}=J_0(E_0/\Omega) J $ compares well to the time averaged results. Main panel: low momentum behavior $k\to 0$, inset: high momentum behavior  $k\to \pi^-$.  }
\label{fig:plot2}
\end{figure}

In Fig.~\ref{fig:plot2} the time averaged results at large times ($Jt>100$) of the momentum dependent correlation function are compared to the ground state prediction  employing the effective (Magnus) Hamiltonian with $J\to J^{\rm eff}=J_0(E_0/\Omega) J $. Very low momentum as well as momentum very close to $\pi$ are difficult to extract faithfully, because we employ a Fourier transform over a finite number of lattice sites, with the maximum of $j=100$. 

The equilibrium properties of the LL phase in our spinless model are characterized by the velocity 
$
v=J\frac{\pi\sin(2\eta)}{2(\pi-2\eta)}\label{v}
$ 
as well as the so-called Luttinger Liquid parameter
$
K=\frac{\pi}{4\eta}\label{K},
$
where $2\eta=\arccos\left(-\frac{U}{J}\right)$. Both $v$ and $K$ depend on the ratio of $U/J$, which is tuned effectively by the drive in the high frequency regime via tuning $J\to J^{\rm eff}=J_0(E_0/\Omega) J $.
A hallmark characteristic of LL behavior can be found in the density-density correlations 
\begin{equation}
C_{nn}(j)=\left\langle (n_0-\frac12)( n_j-\frac12) \right\rangle=\frac{K}{2\pi^2 j^2}+C(-1)^j\left(\frac{1}{j}\right)^{2K},\label{eq:defCnn}
\end{equation}
which sensitively depend on $K$. At low momentum $k\to 0 $ this Fourier transforms
to 
\begin{equation}
C_{nn}(k)\stackrel{k\to 0}{\longrightarrow}\frac{K}{2\pi}k\label{eq:Cnnlowk},
\end{equation}
while at momentum approaching $k\to \pi^-$ the correlations follow a power-law divergence 
\begin{equation}
C_{nn}(k)\stackrel{k\to \pi^-}{\longrightarrow}(\pi-k)^{2K-2}.\label{eq:Cnnhighk}
\end{equation}

We find an unexpected level of agreement of the time averaged results obtained for the periodically driven system and the equilibrium results in the ground state with the appropriate modification $J\to J^{\rm eff}=J_0(E_0/\Omega) J $, even at very fast ramps of the driving downto $\tau=0.1$. This is fascinating as quenches in LLs generically lead to faster entanglement growth and the time-scales accessible here are out of reach of DMRG simulations. This means that surprisingly the  driving protocol maps the initial ground state more faithfully to the ground state of the effective (time-averaged) model than the quench does, even for very fast ramps of the driving protocol. The reduction in entanglement growth (as indicated already by the reachable time-scales) and heating might provide an intriguing shortcut to adiabatic state preparation in LLs. The two physical protocols of fast ramp of periodic driving and the quench are only formally equivalent if $\tau \ll 1/\Omega$. As indicated here this allows to use fast ramps $J\tau=0.1$, without introducing large excitation energy in the picture of the effective Hamiltonian (if the frequency is sufficiently large). 

\section{SI: Tuning a CDW}
\begin{figure}[t]
\centering
\includegraphics[width=\columnwidth]{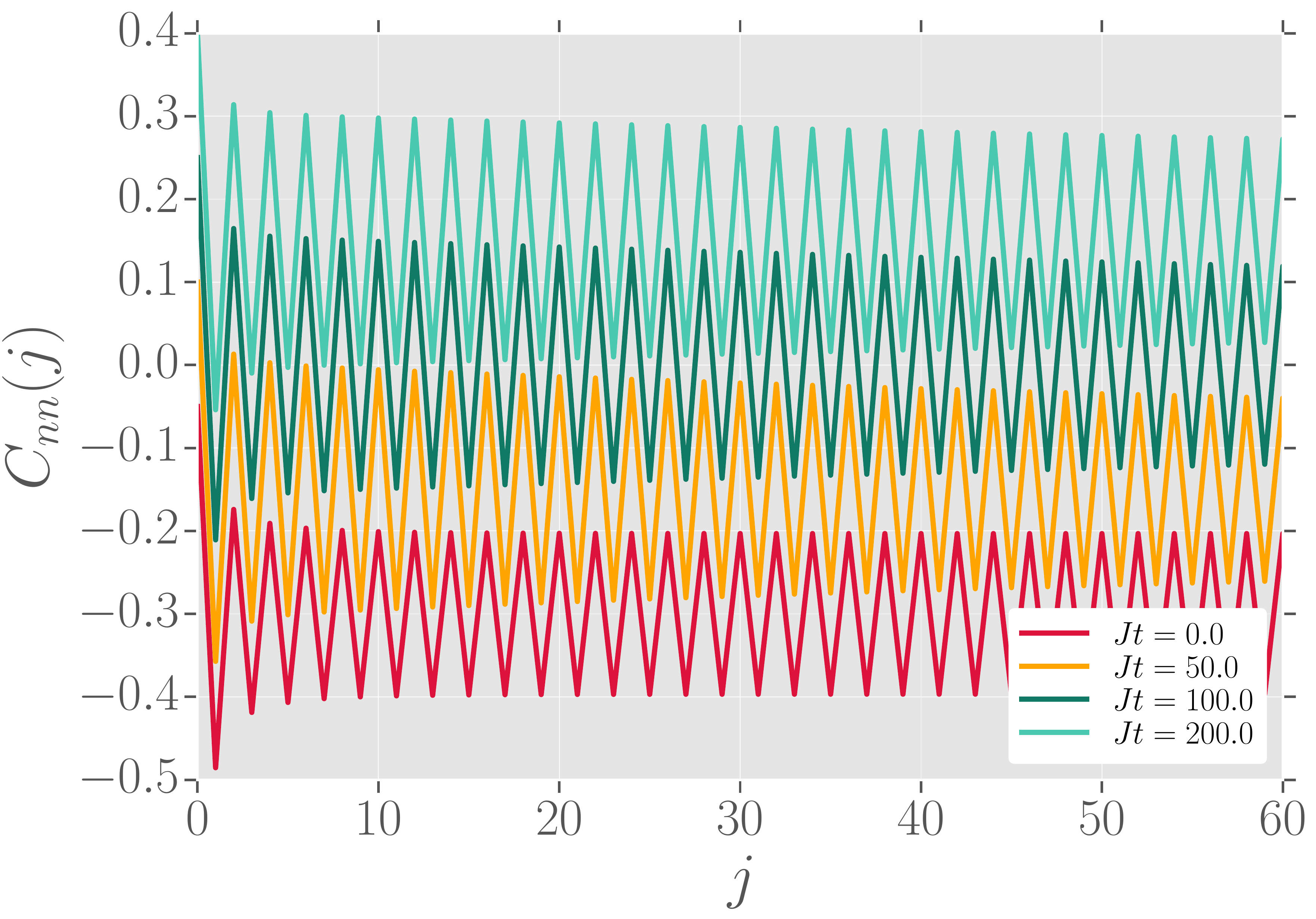}
\caption{ The correlation function $C_{nn}(j)$ at different times averaged over one drive period. Lines are shifted vertically for clarity of depiction; the midpoint of the oscillation is zero.  The parameters are $U/J=1.7$, $E_0/\Omega=1.5$ $J\tau=2$and $T/J=0$.   }
\label{fig:plot_CDW_time_slices}
\end{figure}

In Fig.~\ref{fig:plot_CDW_time_slices} we show the correlation function $C_{nn}(j)$ at different times averaged over one drive period. The curves for $Jt=100$ and $Jt=200$ are stationary on the scale of the plot.

\section{SI: Effective Temperature Fits}
\begin{figure}
\centering
\includegraphics[width=\columnwidth]{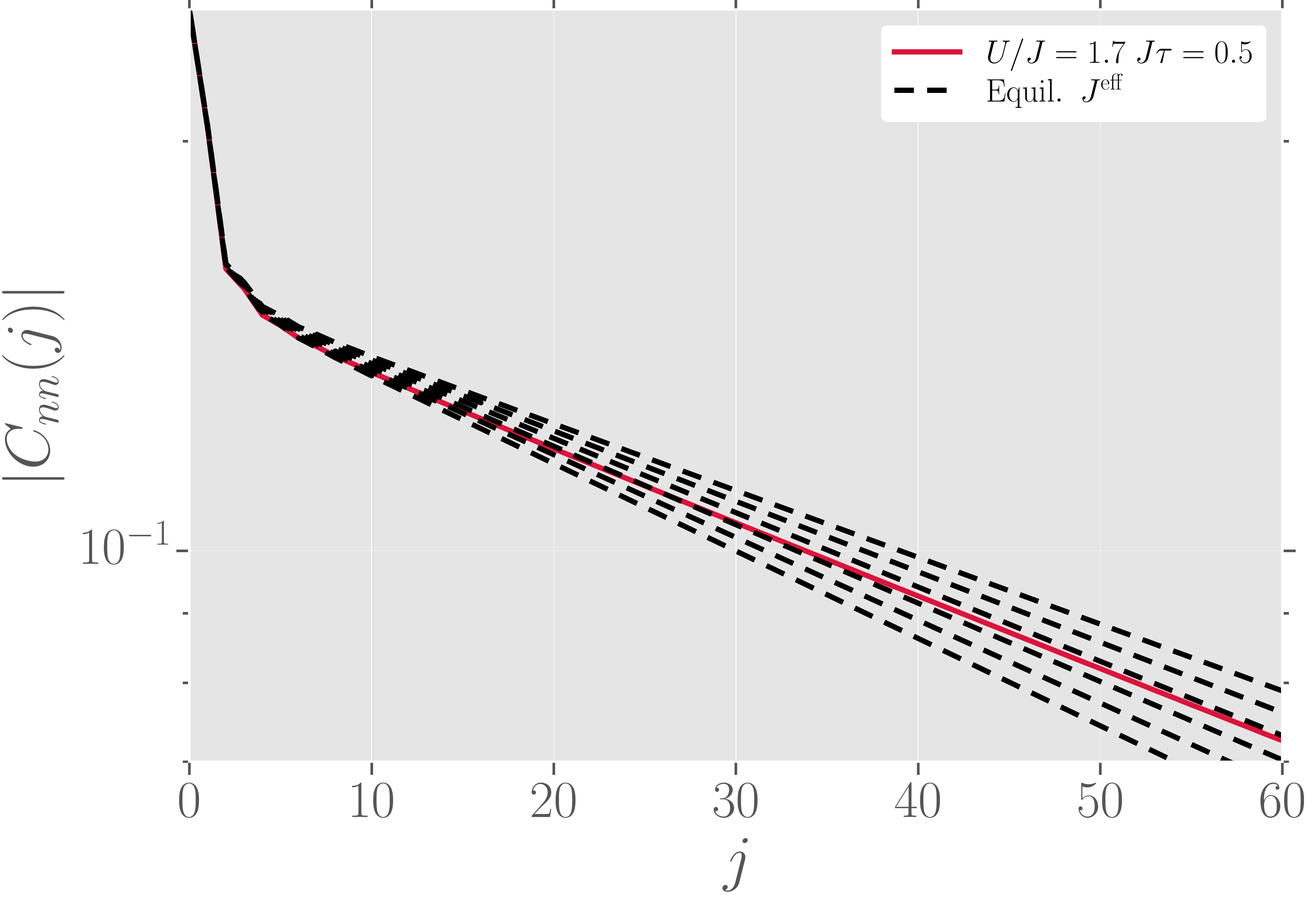}
\caption{Time-averaged density-density correlation function $C_{nn}(j)$ for $U/J=1.7$ and $J\tau=0.5$ (solid line) compared to different thermal expectations obtained in the Magnus formalism. From bottom to top the temperatures of the thermal expectations are $J/T=11,11.2,11.4,11.6,11.8,12.0$. }
\label{fig:plotTfit}
\end{figure}
In the main text we discussed that using a thermal ensemble within the Magnus formalism captures the exponential decay in the density-density correlation function found in the driven system. However, the functional form of these correlations deviate between the driven case and the thermal equilibrium ensemble. To exemplify this point we show one curve obtained for driving the system compared  to the thermal expectations using the Magnus formalism and varying $T$ in Fig.~\ref{fig:plotTfit}. A satisfactory level of agreement can either be obtained at small $j$ but then the slope of exponential decay is not well captured and deviations show up at larger $j$ or the slope of the exponential decay can be fitted well, but then the deviations at small $j$ lead to a parallel offset even at large $j$. 

Next we estimate the form of the effective temperature in this case. Modifying the hopping effectively to smaller values further stabilizes the CDW phase and the gap $\Delta$ as well as the asymptotic value of the correlations $\lim_{j\to\infty}C_{nn}(j)=(-1)^jC_{nn}^\infty$, rise. We show the time averaged real space correlations $C_{nn}(j)$ compared to the equilibrium prediction for $J\to J^{\rm eff}$ in Fig~3 of the main plot. 
The finite speed of the ramp does lead to an, approximately exponential, decay depending on $\tau$ in the long range correlations after the ramp. This decay looks qualitatively  similar to results obtained at finite temperature, for which the rate of the exponential decay of the correlations at large $j$ scales with $T$ (only at $T=0$ one obtains strictly long ranged correlations in equilibrium). Although the decay looks  exponential, we emphasize here, that comparing the correlation function for the entire range of $j$ to thermal expectations using the effective Hamiltonian $J\to J^{\rm eff}$ gives a poor fit (as expected for integrable systems) for any value of $T$, because one can only either fit well the tail with the right exponential decay at large $j$ or fit well the small $j$ behavior (see above). 
In the following, we neglect the details at small $j$ and define an effective temperature using the effective Hamiltonian $J\to J^{\rm eff}$ only by fitting to the asymptotic decay at large $j$ (shown for one example as circles in Fig.~3). This yields a scaling of the effective temperature as shown in the  inset of Fig.~3 for the two values of $U$ shown in the main panel.  At large values of $U$, we estimate the amount of energy injected into the system, by relating the Fourier transform of the ramp  envelope $\sim \tau /\sinh\left[(\pi \tau \omega)/2\right]$ to the gap. If we set the value of the gap to be $\Delta$ (calculated from $J$ and $U$), the energy injected in the system (which is proportional to the temperature) should follow the general form $T^{\rm eff}/\Delta^{\rm eff}= a \Delta\tau\sinh\left(b \tau \Delta\right)$, with $a$ and $b$ unknown proportionality constants. Here $\Delta^{\rm eff}$ and $\Delta$ are determined from $U/J^{\rm eff}$ and $U/J$, respectively. Indeed such a behavior is found to agree well with the numerically extracted effective temperatures for large $U/J=4$ (see dashed line in inset). 

\begin{figure*}[b]
\centering
\includegraphics[width=2\columnwidth]{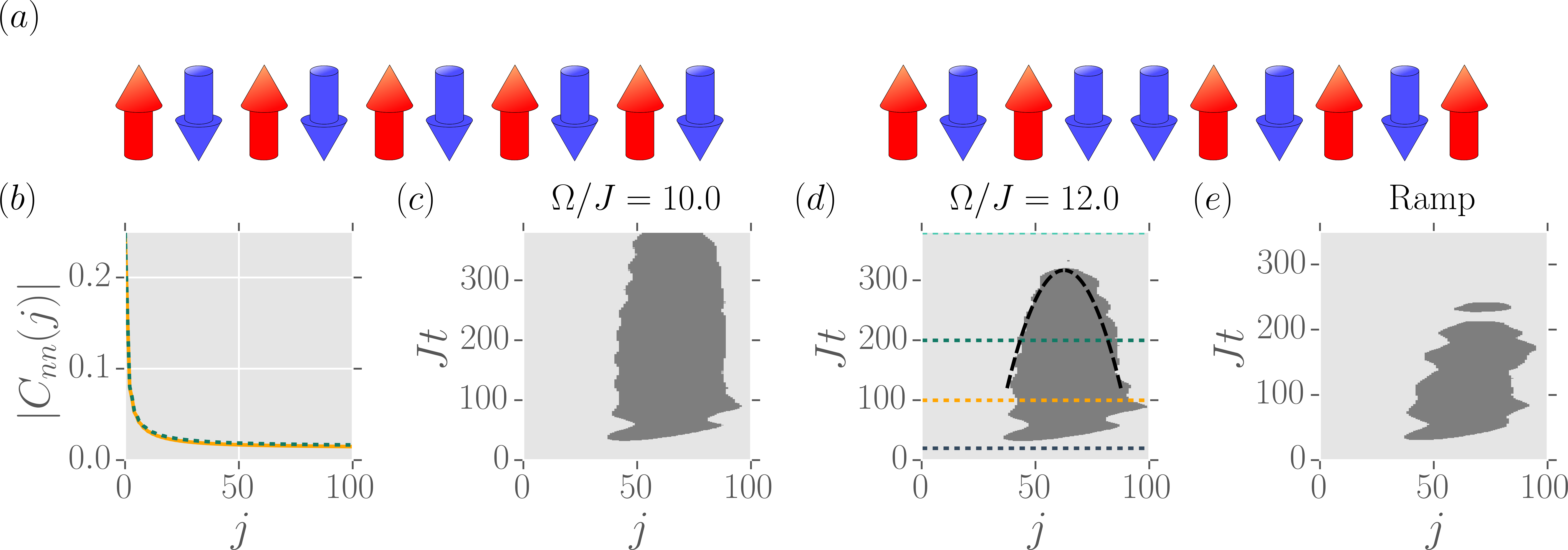}
\caption{(a) Depiction of a phase slip, where positive correlations are depicted by an "up spin" and negative correlations by a "down spin". Left: regular CDW pattern. Right: one phase slip between the forth and fifth "spin". (b) Time-averaged correlation function $C_{nn}(j)$ for $U/J=0.95<1$ and $E_0/\Omega=1$ (solid line) compared to the equilibrium prediction (dashed line) with $J\to J^{\rm eff}=J_0(E_0/\Omega) J $ (such that $U/J^{\rm eff}>1$). The other parameters are $J\tau=4$, $\Omega/J=10$ and $T/J=0$. (c) and (d) time evolution of the phase $\phi(t,j)$ averaged over the micro motion for two values of $\Omega$. Light and dark gray denote the phase of $+1$ and $-1$, respectively. (e) the same as (c) and (d) but instead driving with a ramp of the hopping following $J(t)=J+(J^{\rm eff}-J)\left[\frac12 +\frac12 \tanh\left(\frac{t}{\tau}\right)\right]$. The other parameters in (c)-(e) are $U/J=0.95$, $E_0/\Omega=1.5$ $J\tau=4$and $T/J=0$. In (d) we also show the time evolution predicted by a simple model of the dynamics of the phase slips (see text) and include as horizontal lines of the same color the times at which the correlation function i shown in the main text.
}
\label{fig:plot4}
\end{figure*}

\section{SI: Time Evolution of Phase-Slip}

In this section we analyze the dynamics of phaseslips as introduced when effectively ramping across the quantum critical point in more detail.
Fig~\ref{fig:plot4}(a) exemplifies the concept of a phase slip in the CDW pattern by displaying $+$ and $-$ correlations as up- and down-spins. To highlight the dynamics of phase-slips we introduce the variable $\phi$, which is $1$ if the CDW pattern has sign which follows the regular (equilibrium) pattern and $-1$ if not.
The likelihood $\lambda$ of a phase slip to occur scales with the excitation energy acquired  by the system through the ramp. In a quasi-classical picture the probability to encounter $n$ phase slips by the time one reached the $j$th site is given by a binomial distribution $P_n(j)={{j}\choose{n}}\lambda^j(1-\lambda)^{j-n}$. Therefore, the smaller the excitation energy the further in $j$ the phase slips are expected to occur as $P_1(j)>P_0(j)\Rightarrow j>j_c$, with $j_c=\frac{1-\lambda}{\lambda}$. Indeed for small excitation energies, the phase slips seem to be beyond the window of $j\leq100$ we calculated, see  Fig~\ref{fig:plot4}(b). However, for increasing the amplitude of the drive field (Fig~\ref{fig:plot4}(c-e)) the phase slips occur well within the region of $j\leq100$. Also for smoother ramps (longer $\tau$, less excitation energy, smaller $\lambda$) the phase slip thus should move to larger values of $j$, which is confirmed by our numerics (see supplementary information). A rough estimate relates the position of the first to the second phase slip (phase and anti-phaseslip pair) by a factor of two at small $\lambda$, because the region where one expects a single phase slip to occur is  roughly given by $P_1(j)>P_0(j)$ and simultaneously $P_1(j)>P_2(j)$, which for small values of $\lambda$ yields $2j_c>j>j_c$. Our numerics (compare Fig~\ref{fig:plot4}(c-e)) are roughly consistent with this predictions before the phaseslip -- anti-phaseslip pair starts to attract and eventually annihilates into spin-waves. The attraction of this pair can be studied in a quasi-particle picture, where the phaseslip and anti-phaseslip pair take the role of particles with $\log(r)$ attractive potential. Because we expect the long-time dynamics of these pairs to be overdamped the simple equation of motion of the relative coordinate for the phaseslip pair is given by $\dot r(t)=C/r(t)$, where $C$ is a constant describing the ratio of the prefactor of the attractive potential and the viscosity of the phase slips' dynamics. The resulting dynamics read $r(t)=r_0 \sqrt{2C(t_0-t)}$, which is valid for times smaller  than the time $t_0$, where the pair comes very close to each other and annihilates by emitting spin-waves. This rough estimate of the dynamics of the phase-slip anti-phaseslip pair is in reasonable agreement with the numerics as shown as a long-dashed line in  Fig~\ref{fig:plot4}(d). Additionally, we report that the frequency $\Omega$ can be used to effectively tune the constant $C$ controlling the attraction felt by the phaseslip anti-phaseslip pair, with the limit $\Omega\to \infty$ restoring the result of a ramp in $J$, as expected by a leading order Magnus expansion.

\begin{figure}[h]
\centering
\includegraphics[width=\columnwidth]{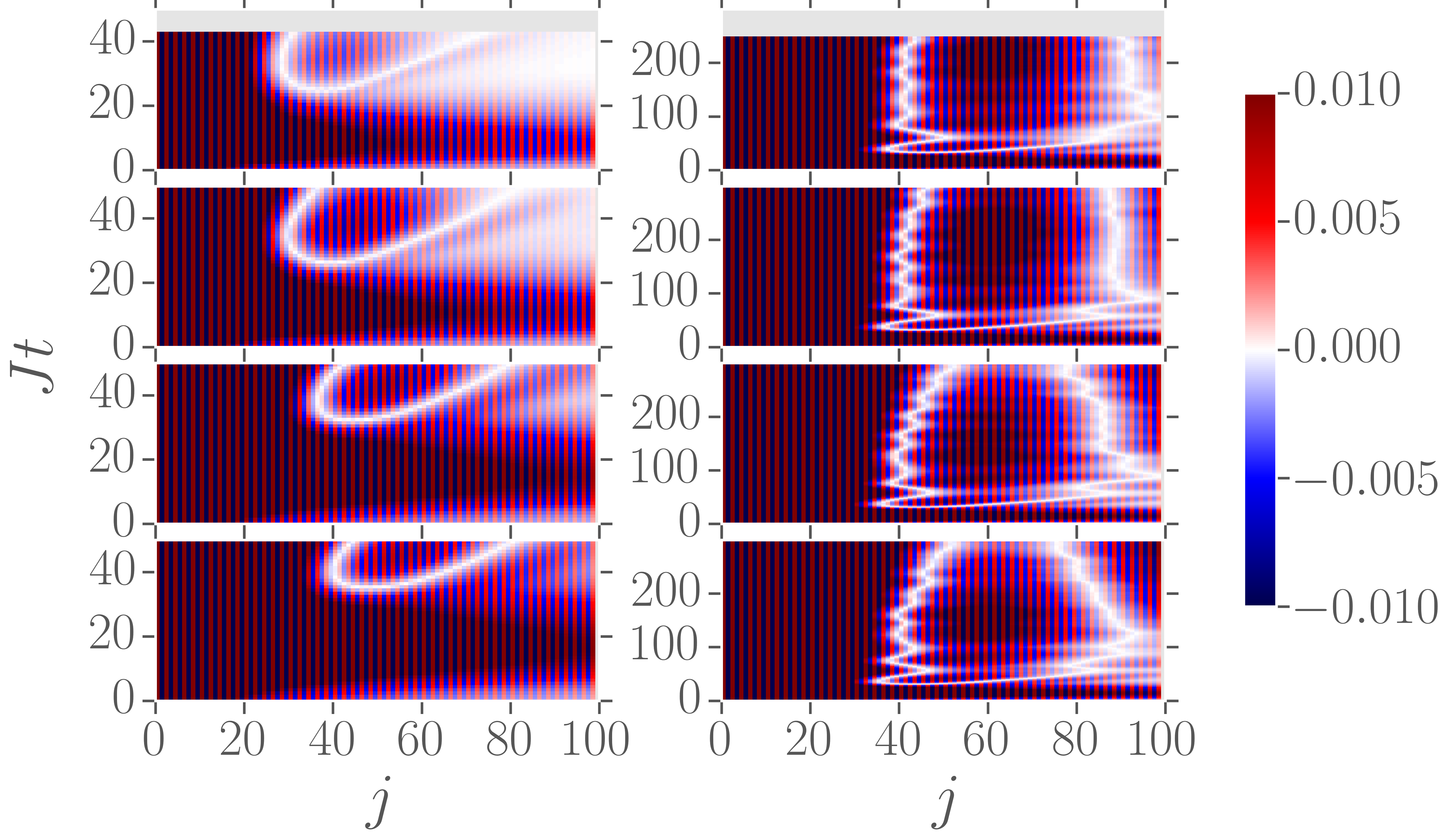}
\caption{False color plot of the  density-density correlations $C_{nn}(j)$ averaged over the micromotion on a diverging color scale (white is zero) which resolves small absolute values. The other parameters are $U/J=0.95$, $E_0/\Omega=1.5$ and $T/J=0$ as well as $\Omega/J=10$ for the left panels and $J\tau=4$ for the right panels. The left panels show $J\tau=1,2,4,5$ and the right panels show $\Omega/J=8,10,12,14$ from top to bottom.   }
\label{fig:plotPS_TS}
\end{figure}

Fig.~\ref{fig:plotPS_TS} shows the density-density correlations $C_{nn}(j)$ averaged over the micromotion as a false color plot utilizing a diverging color scheme (white is zero, to highlight the dynamics of the phase slip). As $\tau$ is increased (less excitation energy is injected into the system), the phaseslip--antiphaseslip pair is created at larger values of $j$ in accordance to the simple model discussed in the main text (compare left panels in Fig.~\ref{fig:plotPS_TS}). Tuning $\Omega$ one can control the time scales on which the phaseslip pair is annealed out, which means one can tune their effective attraction (compare right panels in Fig.~\ref{fig:plotPS_TS}). The weak dependence of the results on $\tau$ is in agreement with the recent study Ref.~\cite{Gardas17}.

\begin{figure}[h]
\includegraphics[width=\columnwidth]{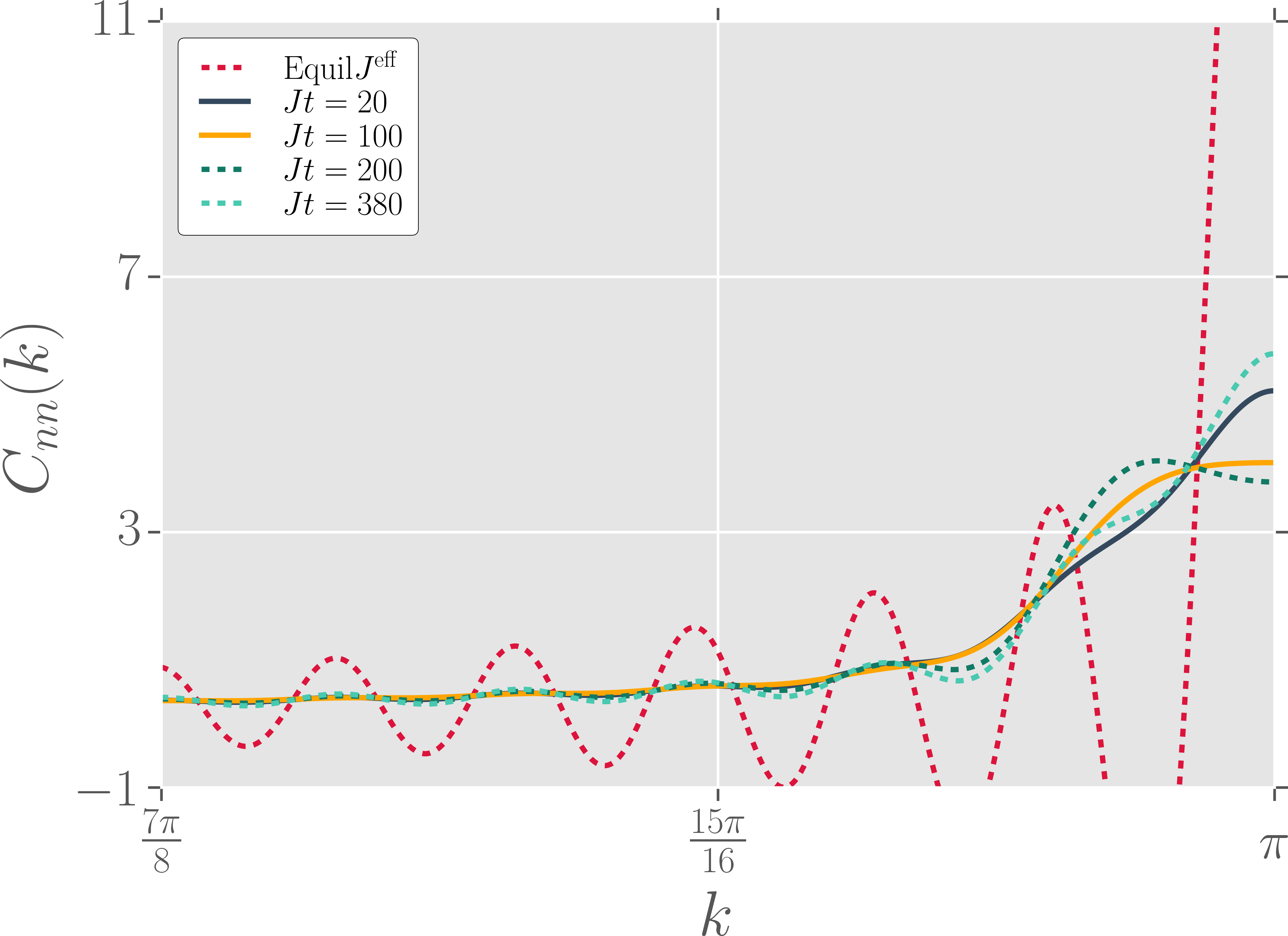}
\caption{Fourier transforms of the $C_{nn}(j)$ shown in Fig~5 as well as the ground state prediction for comparison. The other parameters are $J\tau=4$, $\Omega/J=12$, $U/J=0.95$, $E_0/\Omega=1.5$ and $T/J=0$.  }
\label{fig:plot4_3}
\end{figure}

Fig~\ref{fig:plot4_3} depicts the consequences of the presence of phaseslips for the  Fourier transform $C_{nn}(k)$ of the correlation function $C_{nn}(j)$. We show $C_{nn}(k)$ before, during and after the phaseslip has occurred. The consequences of phase-slips for the Fourier transform $C_{nn}(k)$, which we calculate over a finite region in space $j<100$ are quite severe. The equilibrium $T=0$ expectation using the effective Hamiltonian from a Magnus expansion would show a pronounced singularity at $\pi$, which (for our finite Fourier transform) shows up as a strong feature plus Gibbs ringing at this momentum. In comparison to that the time evolved correlation functions show a strongly suppressed feature, with a maximum that shifts slightly as the phaseslip pair is created. At the time the phaseslips anneal out the maximum shifts back and starts to grow again.

\section{SI: Heating in Regime (b)}

\begin{figure}[t]
\centering
\includegraphics[width=\columnwidth]{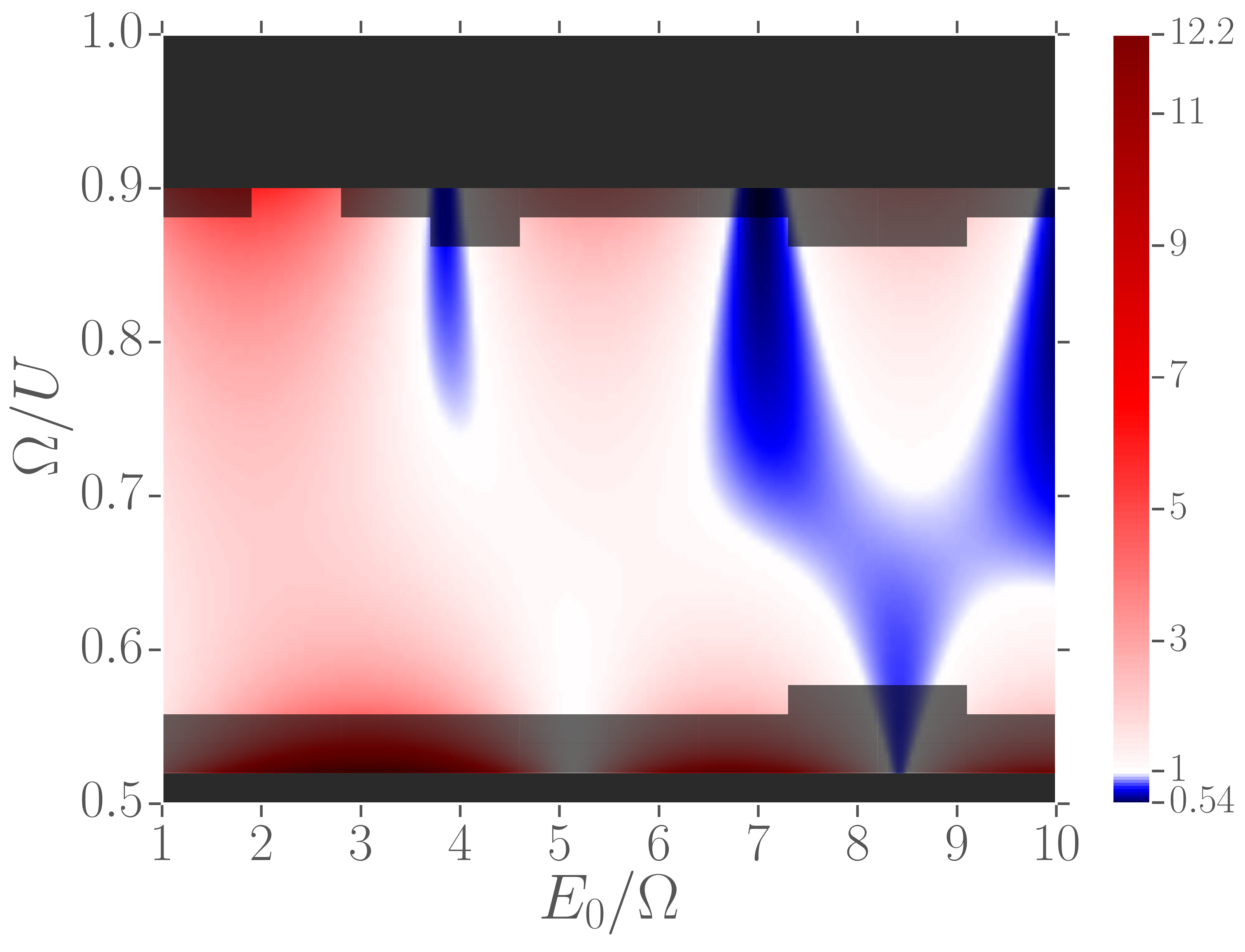}
\caption{False color plot of the  effective hopping $J^{\rm eff}$ following Eq.~(3) using a diverging color scheme (white is one). We block out regions where heating becomes important and large enough time scales to analyze the dynamics cannot be accessed by DMRG, due to entanglement growth. Dark blacked out regions show a much stronger heating rate than semi-transparent blocked out regions.    }
\label{fig:plot_PD.pdf}
\end{figure}

In Fig.~\ref{fig:plot_PD.pdf} we show the effective hopping $J^{\rm eff}$ following Eq.~(3) as a false color plot using a diverging color scheme (white is one). The non-monotonic nature in which $J^{\rm eff}$ can be controlled results in blue and red regions corresponding to effectively increasing or decreasing the hopping by the external drive. We concentrate on $\Omega/U>0.5$ where resonances are avoided and indicate by semi-transparent or solid blacked out  regions the parameter space which leads to heating, and consequently entanglement growth.

\begin{thebibliography}{}

\bibitem{Basov17}
D. N. Basov, R. D. Averitt and D. Hsieh,
Nature Materials {\bf 16}, 1077 (2017).

\bibitem{Mankowsky16rev}
 R. Mankowsky, M. Forst and A. Cavalleri, Reports on Progress in Physics, {\bf 79} 064503 (2016).

%
%
%
%
%
%
%
%
%
%
%
%
%
%
%
%



\bibitem{Shirley65}
J. H. Shirley, Phys. Rev. {\bf 138}, B979 (1965).
 
\bibitem{CohenTannouji98}
C. Cohen-Tannoudji, J. Dupont-Roc, and G. Grynberg, {\it Atom-Photon Interactions}, (Wiley Interscience, Chichester, UK, 1998).


\bibitem{Eckardt05}
A. Eckardt, C. Weiss, and M. Holthaus, Phys. Rev. Lett {\bf 95}, 260404 (2005).

\bibitem{Zenesini09}
A. Zenesini, H. Lignier, D. Ciampini, O. Morsch, and E. Arimondo, Phys. Rev. Lett. {\bf 102}, 100403 (2009).

\bibitem{Itin15}
A. P. Itin and M. I. Katsnelson
Phys. Rev. Lett. {\bf 115}, 075301 (2015).

\bibitem{Meinert16}
F. Meinert, M. J. Mark, K. Lauber, A. J. Daley, and H.-C. N\"agerl
Phys. Rev. Lett. {\bf 116}, 205301 (2016).

\bibitem{Abanin15}
D. A. Abanin, W. De Roeck, and F. Huveneers
Phys. Rev. Lett. {\bf 115}, 256803 (2015).

\bibitem{Mori16}
T. Mori, T. Kuwahara, and K. Saito
Phys. Rev. Lett. {\bf 116}, 120401 (2016).

\bibitem{Kuwahara16}
T. Kuwahara, T. Mori, K. Saito,
Annals of Physics {\bf 367} (2016).


\bibitem{Mentink15}
J.~H.~Mentink, K.~Balzer and M.~Eckstein, Nature Communications {\bf 6}, 6708 (2015). 
\bibitem{Claassen16}
M.~Claassen, H.-C.~Jiang, B.~Moritz and T.~P.~Devereaux,
Nature Communications {\bf 8}, 1192 (2017)  

\bibitem{Peronaci17}
F. Peronaci, M. Schir\'o, O. Parcollet,
arXiv:1711.07889.

\bibitem{Magnus64}
W. Magnus,
Comm. Pure and Appl. Math. {\bf 7}, 649 (1964).

\bibitem{Bukov15}
M. Bukov, L. D'Alessio, A, Polkovnikov
 Adv. Phys. {\bf 64}, 139 (2015).


\bibitem{Bukov16}
M. Bukov, M. Kolodrubetz and A. Polkovnikov
Phys. Rev. Lett. {\bf 116}, 125301 (2016)






 

\bibitem{Poletti11}
D. Poletti and C. Kollath,
Phys. Rev. A {\bf 84}, 013615 (2011)


\bibitem{Mendoza-Arenas17}
J. J. Mendoza-Arenas, F. J. Gomez-Ruiz, M. Eckstein, D. Jaksch, and S. R.  Clark,  Ann. Phys. (Berlin) {\bf 529} 1700024 (2017).

\bibitem{Eurich11}
 N. Eurich, M. Eckstein, and P. Werner
Phys. Rev. B {\bf 83}, 155122 (2011).

\bibitem{Herrmann17}
A. Herrmann, Y. Murakami, M. Eckstein, P. Werner
arXiv:1711.07241.

 \bibitem{Schuler17}	
M. Sch\"uler, Y. Murakami, P. Werner
arXiv:1712.06098.

\bibitem{Weinberg17}
P. Weinberg, M. Bukov, L. D’Alessio, A. Polkovnikov, S. Vajna, M. Kolodrubetz, Physics Reports {\bf 688}  1 (2017).

\bibitem{KZ} 
T.~W.~B.~Kibble, J. Phys. A {\bf 9}, 1387 (1976); Phys. Rep. {\bf 67},
183 (1980); W.~H.~Zurek, Nature (London) {\bf 317}, 505
(1985); Acta Phys. Pol. B {\bf 24}, 1301 (1993); Phys. Rep.
{\bf 276}, 177 (1996).

\bibitem{Manmana09}
S. R. Manmana, S. Wessel, R. M. Noack and A. Muramatsu,
Phys. Rev. B {\bf 79}, 155104 (2009).

\bibitem{SI}
supplementary information
\bibitem{White92} S.~R.~White, Phys. Rev. Lett. {\bf 69}, 2863 (1992).
\bibitem{Vidal07} G.~Vidal, Phys. Rev. Lett. {\bf 98}, 070201 (2007).
\bibitem{Schollwock11} U.~Schollw\"ock, Ann. Phys. {\bf 326}, 96 (2011).
\bibitem{Karrasch12a} C.~Karrasch, J.~E.~Moore, Phys. Rev. B {\bf 86}, 155156 (2012).
\bibitem{Barther13}  T~ Barthel, New J. Phys. {\bf 15}, 073010 (2013).
\bibitem{Kennes16} D.~M.~Kennes and C.~Karrasch, Comput. Phys. Commun. {\bf 200}, 37 (2016).

\bibitem{Lorenzo17}
S. Lorenzo, J. Marino, F. Plastina, G. M. Palma and T. J. G. Apollaro,
Scientific Reports {\bf 7}, 5672 (2017).

\bibitem{Cazalilla06}
M. A. Cazalilla, Phys. Rev. Lett. {\bf 97}, 156403 (2006).

\bibitem{Rentrop12}
J. Rentrop, D Schuricht and V Meden,
New J. Phys. {\bf 14}, 075001 (2012).


\bibitem{Kennes13}
  D. M. Kennes and V. Meden,
  Phys. Rev. B {\bf 88}, 165131 (2013).
  
\bibitem{Mitra13}  
Aditi Mitra,
Phys. Rev. B {\bf 87}, 205109 (2013).
\bibitem{Gardas17} 
B.~Gardas, J.~Dziarmaga, and W.~H.~Zurek,
Phys. Rev. B {\bf 95}, 104306 (2017).


\bibitem{Schlenker85}
C. Schlenker, H. Schwenk, C. Escribe-Filippini, and J. Marcus, Physica B+C {\bf 135}, 511 (1985).
\bibitem{Jerome82}
D. J\'erome and H. Schulz, Adv. Phys. {\bf 31}, 299 (1982).
\bibitem{He99}
H. Haifeng and Z. Dianlin, Phys. Rev. Lett. {\bf 82}, 811 (1999).

\bibitem{Gardas17} 
B.~Gardas, J.~Dziarmaga, and W.~H.~Zurek,
Phys. Rev. B {\bf 95}, 104306 (2017).
\end{thebibliography}
\end{document}